\documentstyle[12pt,epsf,epsfig]{article}
\def\be{\begin{equation}}
\def\ee{\end{equation}}
\def\ba{\begin{eqnarray}}
\def\ea{\end{eqnarray}}
\def\bdm{\begin{displaymath}}
\def\edm{\end{displaymath}}
\def\la{~\mbox{\raisebox{-.6ex}{$\stackrel{<}{\sim}$}}~}

\def\bq{\begin{quote}}
\def\eq{\end{quote}}

 at 10truept

\newcommand{\beq}{\begin{equation}}
\newcommand{\eeq}{\end{equation}}
\newcommand{\beqa}{\begin{eqnarray}}
\newcommand{\eeqa}{\end{eqnarray}}

\def\la{~\mbox{\raisebox{-.6ex}{$\stackrel{<}{\sim}$}}~}

\def\ltap{\ \raise.3ex\hbox{$<$\kern-.75em\lower1ex\hbox{$\sim$}}\ }
\def\gtap{\ \raise.3ex\hbox{$>$\kern-.75em\lower1ex\hbox{$\sim$}}\ }
\def\gl{\ \raise.5ex\hbox{$>$}\kern-.8em\lower.5ex\hbox{$<$}\ }
\def\roughly#1{\raise.3ex\hbox{$#1$\kern-.75em\lower1ex\hbox{$\sim$}}}

\begin{document}

\thispagestyle{empty}
\begin{flushright}
hep-th/0601110\\ January 2006
\end{flushright}
\vspace*{1cm}
\begin{center}
{\Large \bf Exact Black Holes and Gravitational Shockwaves}\\
\vspace*{.2cm} {\Large \bf
on Codimension-2 Branes}\\
\vspace*{.9cm} {\large Nemanja Kaloper\footnote{\tt
kaloper@physics.ucdavis.edu} and
Derrick Kiley\footnote{\tt dtkiley@physics.ucdavis.edu}}\\
\vspace{.5cm} {\em Department of Physics, University of
California, Davis,
CA 95616}\\
\vspace{.15cm} \vspace{1.2cm} ABSTRACT
\end{center}
We derive exact gravitational fields of a black hole and a
relativistic particle stuck on a codimension-2 brane in $D$
dimensions when gravity is ruled by the bulk $D$-dimensional
Einstein-Hilbert action. The black hole is locally the
higher-dimensional Schwarzschild solution, which is threaded by a
tensional brane yielding a deficit angle and includes the first
explicit example of a `small' black hole on a tensional $3$-brane.
The shockwaves allow us to study the large distance limits of
gravity on codimension-2 branes. In an infinite locally flat bulk,
they extinguish as $1/r^{D-4}$, i.e. as $1/r^2$ on a $3$-brane in
$6D$, manifestly displaying the full dimensionality of spacetime.
We check that when we compactify the bulk, this special case
correctly reduces to the $4D$ Aichelburg-Sexl solution at large
distances. Our examples show that gravity does not really obstruct
having general matter stress-energy on codimension-2 branes,
although its mathematical description may be more involved.

\vfill \setcounter{page}{0} \setcounter{footnote}{0}
\newpage

\section{Introduction}

The emergence of the braneworld paradigm has spurred a lot of work
in the exploration of gravity in spaces with defects and/or
boundaries of various codimension. Among the higher-codimension
setups, the codimension-2 branes \cite{raman}-\cite{ghesha} gained
attention because in asymptotically locally flat environs, their
tension curves only the two transverse directions, cusping them
into a cone centered at the location of the brane. This behavior
is modified for different bulk asymptotics \cite{ruth} and for
branes residing on intersections of codimension-1 objects
\cite{ahddk,origami}. The attempts to use this `off-loading' of
the brane vacuum energy into the bulk for alleviating the $4D$
cosmological constant problem \cite{luty,cgn} have been found to
require the usual finely tuned adjustments of parameters once
compactification is enforced to produce $4D$ gravity at large
distances. Indeed, to get an intrinsically flat brane one must
have very particular boundary conditions in the bulk, which
requires adjusting\footnote{Such adjustments are by necessity
global, in spite of the `local guise' as a change of the conical
angle; the change extends to the end of the world in the bulk due
to the peculiarities of `transverse' $2+1~D$ gravity.} the bulk
sector in some way to maintain the brane's flatness upon a change
of matter sector parameters \cite{ruth}, \cite{kmo}-\cite{msled}.

Nevertheless the curiosity that tensional branes can remain
intrinsically flat provoked the study of setups with codimension-2
branes. Surveying the dynamics with a generic stress-energy on a
thin brane in an empty bulk, \cite{cline2} asserted that there is
an inconsistency. They claimed that bulk Einstein's equations
describing codimension-2 branes with $\delta$-function
stress-energy allowed only pure tension $\lambda$, with
$T_{\mu\nu} = - \lambda g_{\mu\nu} \delta^{(2)}(\vec y)$ in
longitudinal directions, and with vanishing transverse components.
Otherwise, noted \cite{cline2}, the solutions would have featured
stronger, non-distributional singularities, that seemed either
unacceptable or difficult to contend with. To handle these
problems frameworks with higher-dimensional operators in the bulk
\cite{bgsn}, thickened, regulated branes
\cite{clinvin}-\cite{tbha}, and combinations thereof were
considered \cite{nasa}. The common goal of these investigations
was to somehow isolate and tame geometric singularities in order
to match geometry and brane stress-energy.

These are all reasonable first-pass strategies, which however
should be pursued carefully since such approaches could be
dangerous, and even deceptive. Because gravity is a theory with a
cutoff, its short distance limits are very tricky. Indeed,
pathologies with distributional sources, similar to those
encountered in codimension-2 setups \cite{cline2}-\cite{nasa} are
already familiar in usual General Relativity (GR). Perhaps the
simplest example arises from the Schwarzschild solution: in the
linearized limit, one may be deceived to think of it as a field of
a $\delta$-function source. In the full theory the short distance
behavior is completely different from the linear theory. When the
exterior geometry is followed inward, at short length scales the
strong nonlinear gravity effects replace the apparent timelike
singularity by a spacelike one, cloaking it with a horizon.
Clearly, we do not throw away the Schwarzschild solution just
because it does not have a $\delta$-function in its core. We
cannot insist on retaining a $\delta$-function source because this
source is itself an {\it approximation}, obtained by
coarse-graining over the interior structure of a realistic lump of
energy. At very short distances, this idealized form will be
modified by corrections from interactions including gravity and
also from quantum mechanics.

Many more examples are provided by line sources in GR
\cite{werner}-\cite{alex}. It is well known that the singularities
in that case are hard to even classify \cite{werner}, and that the
limiting procedures involving distributions that would reproduce
the fields of static straight symmetric $\delta$-function sources
are cumbersome and ambiguous \cite{getra}. Nevertheless, this has
not hampered deriving cosmic string solutions and exploring their
dynamics \cite{alex}. This program revealed that the thin strings
in conical spaces with $\delta$-function stress-energy are really
an idealization, and that in more realistic situations, when local
strings wiggle or when they are perturbed by local inhomogeneities
of matter on them, they will develop long range Newtonian
potentials in transverse directions. Although this may modify the
conical geometry at short and long ranges \cite{alex}, as long as
the asymptotic geometry very far from the disturbance relaxes to a
conical space, they can be viewed as legitimate string
configurations. An extreme case in point are the black holes
pierced by cosmic strings found by Aryal, Ford and Vilenkin (AFV)
\cite{afovi}, where the geometry of the local string asymptotes a
line distribution with a conical deficit far away from the black
hole, but is tremendously deformed near the hole by its strong
nonlinear fields. While it was not immediately clear that this
solution is a limit of some distributional geometry, later on
\cite{acgrk} it was shown how to obtain it by a limiting procedure
in the $4D$ gravitating Abelian Higgs model.

In light of this one may argue \cite{alexcomm} that to understand
the low energy limit of general cosmic strings one ought to look
for physically interesting solutions with conical structure in the
bulk, even if they include some short distance deformations. We
take the view that the example of \cite{afovi,acgrk} is a
concrete, if fortuitous, evidence in favor of \cite{alexcomm}, and
we follow this directive here. This immediately yields an
unexpected prize: the family of exact metrics for a black hole
stuck on a codimension-2 brane. This family of solutions is a
generalization of the $4D$ AFV black hole on a string, and
includes the very first {\it explicit}, exact localized black hole
on a $3$-brane\footnote{There exist black hole solutions on a
2-brane in $4D$ \cite{emhomy} but it is hard to extend them to
higher dimensions. Some interpretations of these difficulties were
offered in \cite{holograms}, and recently some interesting
non-vacuum solutions with $AdS_4 \subset AdS_5$ asymptotics were
studied \cite{ggi}.}, that can be used for computing black hole
production and decay rates at the LHC \cite{bhlhc}. Although we do
not have a realization of these solutions as a limit of some
distributional source interacting with a black hole, we expect
that such a picture should exist, possibly along the lines of the
$4D$ resolution of the AFV solution as in \cite{acgrk}.

Our black holes provide us with a direct clue how to find another
family of solutions with matter sources localized to a thin brane,
where the curvature singularities remain tame even when the matter
stress energy is not pure tension. They are exact gravitational
fields of a relativistic particle stuck on a thin codimension-2
brane in $D$ dimensions, and include the fields of photons living
on a $3$-brane in a $6D$ flat spacetime. Such solutions can be
understood as a brane black hole boosted to a relativistic speed,
in a way analogous to the Aichelburg-Sexl solution of GR
\cite{aichsexl}, and just like it carrying only a
$\delta$-function singularity along its worldline. To obtain the
shockwaves, we employ the cutting and pasting techniques of
\cite{thooft,kostas} which have already been applied to braneworld
models in \cite{roberto,kallet,kaso}, rather than directly
boosting the black hole. It turns out that our shockwaves look
just like the higher-dimensional shocks \cite{higherdwaves}, which
however live on a conical singularity in the bulk, instead of a
flat background. Specifically in the case of a 3-brane in $6D$
they depend on the transverse distance from the source as $1/r^2$.
To see how to recover $4D$ long range gravity in this case, we
close off the bulk by imposing periodic boundary conditions for
bulk fields, as a toy model of compactification. The shockwaves
then correctly reproduce the $4D$ Aichelburg-Sexl solution at
distances larger than the period of compactification, whose
long-range fields vary as a logarithm of the transverse distance
from the source. Examining the black hole and its shockwave limit,
we elucidate the short-distance scales at which the nonlinear
effects of the gravitational fields of brane-localized objects
start to distort the bulk, which should be useful in the search
for regulated versions of codimension-2 braneworlds with matter.
This supports our view, motivated by \cite{alexcomm}, that gravity
by itself {does not} really obstruct having localized sources on
codimension-2 branes, but may merely obscure the way we see them.

\section{Field Equations and Vacua}

We start with a brief review of the field equations and vacuum
solutions describing tensional straight codimension-2 branes in
$D$ dimensions. We assume that gravity propagates in the bulk as
governed by the standard $D$-dimensional Einstein-Hilbert action.
We further assume that the stress-energy sources are completely
localized to a codimension-2 object, vanishing elsewhere in the
bulk. This allows us to seek metrics of the form
\be ds^2 = {\cal F}^2(y) g_{\mu\nu}(x) dx^\mu dx^\nu + h_{ab}(y)
dy^a dy^b \, , \label{eqn:metrics} \ee
where the brane is located at the center of the bulk at $y^a=0$,
and is at rest. The field equations in an empty bulk with a brane
and a brane-localized stress energy tensor $T^\mu{}_\nu$ are
\be M_D^{D-2}G^A{}_{B} = T^\mu{}_\nu \delta^A{}_\mu \delta_B{}^\nu
\frac{1}{\sqrt{\det h}}\delta^{(2)}(y) \, ,\\
\label{eqn:ddimeom} \ee
where the coordinates $x^\mu$, $\mu \in \{0, \ldots, D-3\}$ cover
the brane worldvolume and the coordinates $y^a$, $a \in
\{D-2,D-1\}$ parameterize the two dimensions transverse to the
brane, while the capital latin indices count over all $D$
coordinates. With the metrization (\ref{eqn:metrics}), the factor
${1}/{\sqrt{\det h}}$ properly covariantizes the tensor density
$\delta^{(2)}(y)$. Here $M_D$ is the bulk Planck mass and
$G^A{}_B$ the bulk Einstein tensor, computed from the full metric
(\ref{eqn:metrics}). The induced metric on the brane, from
(\ref{eqn:metrics}), is ${\cal F}^2(0) g_{\mu\nu}$, and as long as
${\cal F}(0)$ is finite we can choose ${\cal F}(0) = 1$ by a
rescaling of transverse coordinates $x^\mu \rightarrow x^\mu/{\cal
F}(0)$. Clearly, if ${\cal F}$ diverges as we approach the brane
at $y^a=0$ things may not be so simple. We will keep this in mind
in what follows. Also, in general we could have introduced the
cross-terms $g_{a\mu}$ in the metric (\ref{eqn:metrics}), for
example by substituting $dy^a \rightarrow dy^a + A^a{}_\mu
dx^\mu$. However from this expression it is clear that in the
brane worldvolume theory such objects would behave as towers of
Kaluza-Klein (KK) vector fields. In what follows we will restrict
our attention to the sector where they vanish, assuming that brane
sources do not carry KK vector charges. More general solutions
with the vectors turned on exist, but are not needed for our
purposes here (see \cite{raman}).

Tracing (\ref{eqn:ddimeom}), we obtain
\be M_D^{D-2} R = - \frac{2T}{D-2} \frac{1}{\sqrt{\det
h}}\delta^{(2)}(y) \, , \label{eqn:ricci} \ee
where $T = T^\mu{}_\mu$, and using this we can break up
(\ref{eqn:ddimeom}) into formulas for the transverse and
longitudinal Ricci tensor components with respect to the brane
worldvolume:
\ba M_D^{D-2} R^{a}{}_b &=& - \frac{T}{D-2} \delta^a{}_b
\frac{1}{\sqrt{\det
h}}\delta^{(2)}(y) \, , \label{eqn:compeomstr} \\
M_D^{D-2} R^\mu{}_{\nu} &=& \Bigl( T^\mu{}_\nu - \frac{1}{D-2} T
\delta^\mu{}_\nu \Bigr) \frac{1}{\sqrt{\det h}}\delta^{(2)}(y) \,
, \label{eqn:compeomslong} \ea
alongside the vanishing cross-terms, $R^a{}_\mu=0$. Note that the
source on the RHS of the longitudinal equation
(\ref{eqn:compeomslong}) is {\it traceless}: indeed, if we split
the brane stress-energy as the sum of the tension, representing
the vacuum energy of the brane matter, and the finite wavelength
matter contributions, $T^\mu{}_\nu = - \lambda \delta^\mu{}_\nu +
\tau^\mu{}_\nu$ respectively, we immediately see that the tension,
being a part of the trace of $T$, immediately cancels from the RHS
of (\ref{eqn:compeomslong}). The longitudinal Ricci tensor
components are only sourced by $T^\mu{}_\nu -  T
\delta^\mu{}_\nu/(D-2) = \tau^\mu{}_\nu - \tau
\delta^\mu{}_\nu/(D-2)$ for any tension $\lambda$. On the other
hand, the sources for the transverse components of the Ricci
tensor $R^a{}_b$ always depend on $\lambda$ explicitly. This
feature of the field equations (\ref{eqn:ddimeom}),
(\ref{eqn:compeomstr}), (\ref{eqn:compeomslong}) is the key
ingredient of the magic of `off-loading' brane vacuum energy into
the bulk \cite{raman,cgn}. While it does not guarantee that the
induced metric on the brane will always be independent of
$\lambda$, it does point out that at large distances along the
brane the induced metric should be essentially independent of
$\lambda$ when the bulk is asymptotically locally Minkowski.

Interesting solutions of (\ref{eqn:ddimeom}) which illustrate this
desensitization of the induced geometry from $\lambda$ are easy to
find. Suppose that the matter stress-energy vanishes,
$\tau^\mu{}_\nu=0$. Since the Ricci tensor is identically zero
away from the brane, the field equations (\ref{eqn:ddimeom}) admit
$D$-dimensional flat space vacuum with the Minkowski metric as a
solution. We factorize the spacetime as a direct product of a
$D-2$-dimensional Minkowski and a $2D$ locally Euclidean and find
that the continuity everywhere away from the brane ensures that
the warp factor is ${\cal F}^2 = 1$ identically. Thus the metric
is exactly
\be ds^2 = \eta_{\mu\nu} dx^\mu dx^\nu + \delta_{ab}(y) dy^a dy^b
\, ,  \label{eqn:minkowski} \ee
where the $2D$ metric $\delta_{ab}(y) dy^a dy^b$ is locally
Euclidean, but the domain of its definition has to be picked in
order to satisfy the equation (\ref{eqn:compeomstr}) on the brane
at $y^a=0$ as well as away from it. A complete cover of the
transverse space is provided by polar coordinates, in terms of
which the metric becomes
\be ds^2 = \eta_{\mu\nu} dx^\mu dx^\nu + d\rho^2 + B^2 \rho^2
d\phi^2 \, , \label{eqn:minkowskipol} \ee
where $\phi \in [0,2\pi]$ and $B$ is picked to solve
(\ref{eqn:compeomstr}). This yields \cite{calc}
\be B = 1 - \frac{\lambda}{2\pi M_D^{D-2}} \, . \label{eqn:ddimB}
\ee
Thus the longitudinal space covered by the coordinates $\rho,
\phi$ is a cone. It is obtained from the flat disk that would
solve (\ref{eqn:ddimeom}) in the $\lambda=0$ limit by extricating
a wedge of angular opening $2\pi(1-B)= \lambda/M_D^{D-2}$,
identifying the edges of the cut and rescaling the polar angle
according to $\phi \rightarrow B \phi$ \cite{raman}. The induced
metric on the brane remains flat, $\propto \eta_{\mu\nu}$, despite
the fact that the brane carried vacuum energy density $\lambda \ne
0$.

It should be clear from this that even if the vacuum brane
(\ref{eqn:minkowskipol}) is perturbed by a localized matter source
described by a $\tau^\mu{}_\nu \ne 0$ of compact support, the long
distance brane geometry may still remain essentially independent
of $\lambda$ as long as the brane straightens out far from the
perturbation. Namely, the metric will receive dramatic {\it
gravitational} corrections (at the very least) near the matter
source, changing its short distance behavior. Such gravitational
short distance corrections should be expected (and were already
pointed at in \cite{cline2,bgsn}): those effects reflect the
nonlinear structure of gravity, accounting for spacetime
distortions as do the strong fields near black holes. However if
the brane straightens out far from the brane matter perturbations,
the bulk geometry far from the brane will converge to the conical
Minkowski form where the deficit angle eats up the tension. One
may expect that the convergence of the bulk geometry to the vacuum
form of (\ref{eqn:minkowskipol}) is rapid, by using the Birkhoff
theorem in higher-dimensional gravity and accounting for the
deficit angle by appropriately renormalizing Newton's constant.
Indeed, if we assume that a regulated perturbed brane exists, then
far from the perturbation the field should converge to that of a
point mass with a deficit angle. In $D$ dimensions, the
gravitational potential of such an object would fall off as
$1/r^{D-3}$, where $r$ is the radial distance away from it, and
hence the geometry should rapidly return to that of
(\ref{eqn:minkowskipol}). The difficulties with this description
should get serious only close in, when nonlinear effects cannot be
disregarded. Thus the scale where the corrections kick up should
be on the order of the gravitational radius of the matter
perturbation. In the next section, we will confirm this intuition
by deriving the exact black hole on a codimension-2 brane, and
determining its gravitational radius $r_0$.

\section{Black Holes Threaded by Codimension-2 Branes}

It is clear from field equations (\ref{eqn:ddimeom}),
(\ref{eqn:compeomstr}), (\ref{eqn:compeomslong}) that away from
the brane the $D$-dimensional Schwarzschild metric,
\be ds^2_D = -\bigg(1-(\frac{r_0}{r})^{D-3} \bigg)
dt^2+\frac{dr^2}{1-(\frac{r_0}{r})^{D-3}} + r^2 d\Omega_{D-2} \, ,
\label{eqn:bhmetric} \ee
remains a solution. Here $r_0$ is the size of the black hole
horizon, determined by its mass, and $d\Omega_{D-2}$ a line
element on a unit $D-2$ sphere $S^{D-2}$. The question is, how is
the black hole solution altered in the presence of the brane. In
general, even for thin branes whose stress-energy tensor may be
imagined to be ultralocal, the presence of the brane may affect
dramatically the black hole horizon, and render the explicit
determination of the geometry describing a black hole on a brane
extremely hard \cite{emhomy,holograms}.

However, this problem greatly simplifies in the codimension-2
case. To illustrate why, let us first discuss a black hole on a
string, given by the AFV solution \cite{afovi}. Finding solutions
of Einstein's equations for a combined gravitational field of some
distribution of matter threaded by a string is very easy if the
matter distribution has an axial symmetry. In this case, all one
needs to do is to orient the string along the axis of symmetry,
and account for its presence by cutting a wedge out of the polar
variable $\phi$, which runs around the symmetry axis. In this way,
one obtains the solution whose geometry at infinity approaches the
conical space of the string, and close in it gets modified by the
gravity of the lump of matter \cite{alex,afovi}. The AFV black
hole is an extreme example of this trick. One simply starts with
the $4D$ Schwarzschild solution, picks the axis, say, in the
North-South direction, along the rays $\theta=0,\pi$ of the $S^2$
transverse to the worldline, and replaces the usual $S^2$ line
element by $d\Omega_2 = d\theta^2 + B^2 \sin^2\theta d\phi^2$,
choosing $B$ to still satisfy Eq. (\ref{eqn:ddimB}) as in the
absence of the black hole. Then the Gauss-Bonnet theorem
guarantees that the full geometry has the same deficit angle as
the string, $2\pi(1-B)$. One can quickly see that this must be the
case because far from the hole, $ds^2_4 \rightarrow -dt^2 + dr^2 +
r^2(d\theta^2 + B^2 \sin^2\theta d\phi^2)$. Upon substituting $z =
r \cos \theta$, $\rho = r \sin \theta$ this can be rewritten in
cylindrical coordinates as $ds^2_4 \rightarrow -dt^2 + dz^2 +
d\rho^2 + B^2 \rho^2 d\phi^2$, i.e. precisely a locally flat
metric with a conical singularity.

We use exactly the same trick to write down the solution
describing a black hole on a codimension-2 brane in $D$
dimensions. This works because, as we have discussed in the
previous section, the field of any thin codimension-2 brane in $D$
dimensions is given by the locally flat metric with a conical
singularity. Thus we can just take the higher-dimensional
Schwarzschild solution, pick an axis and thread a codimension-2
brane along the axis by cutting out a wedge from the range of the
polar angle around this axis, with the opening adjusted to match
the tension of the brane according to (\ref{eqn:ddimB}).

This `brane surgery' is most easily performed when we start with
the black hole solution in uniform coordinates, in terms of which
the metric is of the form $ds^2_D = -F dt^2 + G d\vec x_{D-1}^2$.
It is straightforward to put the solution (\ref{eqn:bhmetric}) in
this form. We replace the radial variable $r$ by ${\cal R}$
according to
\be r =  {\cal R} \, \Bigl( 1 + \frac{r_0^{D-3}}{4{\cal
R}^{D-3}}\Bigr)^{\frac{2}{D-3}} \, ,\label{eqn:rofR} \ee
which yields
\be ds^2_D = - \Bigl(\frac{4 {\cal R}^{D-3} - {r_0}^{D-3}}{4 {\cal
R}^{D-3} + {r_0}^{D-3}} \Bigr)^2 \, dt^2 \, + \, \Bigl(1+ \frac14
(\frac{{r_0}}{{\cal R}})^{D-3} \Bigr)^{\frac{4}{D-3}} \,
\Bigl(d{\cal R}^2 + {\cal R}^2 d\Omega_{D-2} \Bigr) \, ,
\label{eqn:bhmetuni} \ee
with conformally flat spatial slices. Next we pick a
$D-3$-dimensional spatial hypersurface of symmetry (as opposed to
merely an axis of symmetry in the $4D$ AFV case), and transform to
cylindrical polar coordinates defined by it, such that $\vec x$ are
coordinates along this hypersurface, and we coordinatize the two
transverse directions by the transverse distance $\rho$ and the
polar angle $\phi$. With these coordinates, we have ${\cal R}^2 =
\vec x^2 + \rho^2$ and $d{\cal R}^2 + {\cal R}^2 d\Omega_{D-2} =
d\vec x^2 + d\rho^2 + \rho^2 d\phi^2$. Finally, to thread in a
codimension-2 brane with tension $\lambda$, we cut a radial wedge in
the $\rho, \phi$ plane of opening $2\pi (1-B) = \lambda/M_D^{D-2}$,
according to Eq. (\ref{eqn:ddimB}), identify the edges, and rescale
the angle $\phi$ to $\phi \rightarrow B \phi$, so that after
rescaling its range is restored to the interval $[0,2\pi)$. Our
final metric is therefore
\be ds^2_D = - \Bigl(\frac{4 ({\vec x^2 + \rho^2})^{\frac{D-3}{2}}
- {r_0}^{D-3}}{4 ({\vec x^2 + \rho^2})^{\frac{D-3}{2}} +
{r_0}^{D-3}} \Bigr)^2 \, dt^2 \, + \, \Bigl(1+ \frac14
(\frac{{r_0}^2}{\vec x^2 + \rho^2})^{\frac{D-3}{2}}
\Bigr)^{\frac{4}{D-3}} \, \Bigl(d\vec x^2 + d\rho^2 + B^2 \rho^2
d\phi^2 \Bigr) \, , \label{eqn:bhmetbrn} \ee
and it represents a black hole, of horizon size $r_0$, stuck on a
codimension-2 brane. In fact, we should note that it is
straightforward to go back to the spherical polar coordinates for
the metric (\ref{eqn:bhmetbrn}) with the brane included. All we
would do is basically return to the Schwarzschild metric
(\ref{eqn:bhmetric}), but with the line element $d\Omega_{D-2}$ on
the unit sphere $S^{D-2}$ replaced by the line element
$d\ell^2_{D-2} = d\Omega_{D-3} + B^2 \prod_{k=1}^{D-3}
\sin^2(\theta_k) d\phi^2$, which is the metric on a unit
$D-2$-dimensional sphere but with a wedge of opening $2\pi(1-B)$
removed from the polar angle $\phi$. This means that the spatial
surfaces of constant radius are topologically spheres, pinched on
the brane by the tension-induced deficit angle. We should also
note that among the black hole solutions (\ref{eqn:bhmetbrn})
probably the most phenomenologically interesting one is $D=6$,
where our solution models an exact small $6D$ black hole residing
on a 3-brane in two extra dimensions,
\be ds^2_6 = - \Bigl(\frac{4 ({\vec x^2 + \rho^2})^{{3}/{2}} -
{r_0}^{3}}{4 ({\vec x^2 + \rho^2})^{{3}/{2}} + {r_0}^{3}} \Bigr)^2
\, dt^2 \, + \, \Bigl(1+ \frac14 (\frac{{r_0}^2}{\vec x^2 +
\rho^2})^{{3}/{2}} \Bigr)^{{4}/{3}} \, \Bigl(d\vec x^2 + d\rho^2 +
B^2 \rho^2 d\phi^2 \Bigr) \, , \label{eqn:bhmetbrn6d} \ee
which can be used for precise and explicit calculations of
production and evaporation of quantum black holes at the LHC, as
in the studies of \cite{bhlhc}.

Let us (very!) briefly review some of the properties of the black
hole family (\ref{eqn:bhmetbrn}). As in the case of the AFV
solution \cite{afovi}, the horizon distance $r_0$ is an
integration constant in (\ref{eqn:bhmetbrn}), and as such
independent of brane tension. So for a fixed $r_0$ the surface
gravity and the Hawking temperature of the hole are completely
independent of the brane. The Euclideanized version of the
solution (\ref{eqn:bhmetric}) then readily yields that the Hawking
temperature, defined by the period of the Euclidean time, is $T_H
= \frac{D-3}{4\pi r_0}$. However, the presence of the brane alters
the relation between the horizon size and the mass of the black
hole controlling its inertia, as measured by the hole's momentum
integrals at asymptotic infinity. More formally, the formula for
the ADM mass of the black hole is corrected because of the deficit
angle. To see this, we can look at the linearized form of the hole
metric (\ref{eqn:bhmetbrn}), which, using spherical polar
coordinates, $d\Omega_{D-2} \rightarrow d\ell^2_{D-2}$, is $ds^2_D
= -\bigg(1-({r_0}/{\cal R})^{D-3} \bigg) dt^2+ \bigg( 1+
\frac{1}{D-3}({r_0}/{\cal R})^{D-3}\bigg) (d{\cal R}^2 + {\cal
R}^2 d\ell^2_{D-2})$. So the mass of the hole is \cite{rob}
\be m = \frac{D-2}{2} M_D^{D-2} r_0^{D-3} \int_{\tt angles}
d\ell_{D-2} \, . \label{eqn:mass} \ee
Since angles run over a $D-2$-dimensional sphere with a deficit
angle, the integral is given by $\Omega_{D-2} B$, where
$\Omega_{D-2} = 2\pi^{\frac{D-2}{2}}/\Gamma(\frac{D-2}{2})$ is the
volume of a unit $S^{D-2}$ and $B$ is the deficit angle parameter in
(\ref{eqn:ddimB}). Introducing a shorthand $\alpha_D = (D-2)
\Omega_{D-2}/2$ for the fixed dimensionless quantities, the mass
is\footnote{Note that the Bekenstein-Hawking entropy $\sim$ area law
for black holes, $S \sim A/G_N$, is properly upheld. Plugging in
this equation the area, $A \sim r_0^{D-2}$, and the coupling on the
cone, $G_N \sim 1/(M_D^{D-2} B)$, we find that $S \sim (r_0
M_D)^{D-2} B$, and so $T_H S \sim S/r_0 \sim r_0^{D-3} M_D^{D-2} B$,
or therefore $T_H S \sim m$ (using Eq. (\ref{eqn:massfin})).}
\be m = \alpha_D \, M_D^{D-2} B \, r_0^{D-3} \, .
\label{eqn:massfin} \ee
Inverting, we find that the horizon size $r_0$ is expressed in
terms of the ADM mass $m$ according to $r_0 =
m^{1/(D-3)}/(\alpha_D M_D^{D-2} B)^{1/(D-3)}$, or, using
(\ref{eqn:ddimB}),
\be r_0 = \Bigl( \frac{2\pi}{2\pi M_D^{D-2} - \lambda}
\Bigr)^{\frac{1}{D-3}} \,
\Bigl(\frac{m}{\alpha_D}\Bigr)^{\frac{1}{D-3}}
 \, . \label{eqn:horizon} \ee
Now, it is clear from the black hole solution (\ref{eqn:bhmetbrn})
and its linearized form that the strong gravity effects and
nonlinear corrections begin to affect the geometry at distances of
the order of $r_0$ from the hole. Because of the equivalence
principle, however, this will remain true even for sources which
have not yet collapsed, but may be stabilized by some matter
interactions. From formula (\ref{eqn:horizon}) it is clear that
the actual scale where this happens depends not only on the mass
sourcing the field, but also on the tension of the brane. For a
fixed value of mass, nonlinear gravity effects could start at
distances much greater than a naive estimate of the gravitational
radius based on a `braneless' higher dimensional gravity, $\propto
M_D^{-1} (m/M_D)^{1/(D-3)}$, because of the conical enhancement of
the gravitational force, as is manifest in (\ref{eqn:horizon}).
The closer the tension is to the bulk scale, which would be
expected by naturalness, and needed to avoid a large $4D$ vacuum
energy upon compactification \cite{ahhsw}, the larger the
gravitational radius of the mass $m$! In effect, the codimension-2
brane behaves as a lightning rod for gravity: the deficit angle
lessens the bulk volume near the brane, which hampers dilution of
gravitational force with distance. Note however that we confirm
the intuition that a version of the higher-dimensional Birkhoff
theorem still applies for masses on a codimension-2 brane, despite
the presence of the brane. The potential drops off as claimed,
according to $1/r^{D-3}$, although the scale beyond which the
nonlinear effects are negligible may be pushed out to distances $
\gg M_D^{-1} (m/M_D)^{1/(D-3)}$. In practice, this implies that in
the attempt to regulate the brane in order to deal with the
effects of strong gravity of some object of mass $m$ as in
\cite{cline2}-\cite{nasa}, one must thicken up the brane to exceed
the gravitational radius of the mass $m$, as given by
(\ref{eqn:horizon}), in order to be able to treat gravity
perturbatively, and depending on the brane tension this scale
could be very large. Gravitational shockwaves, which we turn to
next, provide us with further examples of this gravitational
lightning rod phenomenon.

\section{Gravitational Shockwaves}

As we have seen above, the nonlinear gravitational corrections at
short distances cannot be neglected any more at scales comparable
to the gravitational radius $r_0$ of the source. Although this
distance may depend on the mass in a complicated way
(\ref{eqn:horizon}) because of the environmental effects, it
really comes about because the mass of the source breaks the
conformal symmetry of the background. Clearly, the smaller the
mass, the shorter the scale where gravitational nonlinearities
become large. This immediately points how to regain some level of
mathematical control over the nonlinearities in the theory, while
continuing to explore nontrivial sectors of gravity. The trick is
to try to suppress the scale at which conformal symmetry is
broken, while keeping a nontrivial stress-energy source to
generate gravity. Clearly, restoring conformal symmetry means
looking at sources whose stress-energy has negligible or vanishing
trace. Hence we should look at the gravitational fields of very
fast particles on the brane. Their gravitational field will be
sourced by the momentum, and the distance scale below which the
nonlinearities are significant will be arbitrarily short,
controlled by the ratio of the rest mass to the momentum of the
particle. In the ultrarelativistic limit, when the rest mass
vanishes, we would expect that the linearized gravity description
would remain valid down to extremely short distances, in which
case we should be able to retain the thin-brane description of
relativistic stress-energy as a $\delta$-function source. Indeed,
this is precisely how the gravitational shockwave solutions work
in conventional GR and in the theories with branes
\cite{aichsexl}-\cite{kaso}. The relativistic limit suppresses the
scale where nonlinearities kick in by restoring the conformal
symmetry of the matter sector, which in turn allows a linear
description all the way to arbitrarily short distances.

To confirm this intuition, we construct the explicit form of the
gravitational shockwaves, sourced by relativistic particles, such
as a photon, on a codimension-2 brane. To do so, we could have
followed the road Aichelburg and Sexl set out on in their seminal
paper \cite{aichsexl}: take our black hole (\ref{eqn:bhmetbrn}),
linearize it, and boost it until its worldline becomes null (but
ensure that in this process we properly gauge-fix the linearized
solution so that no non-physical divergences are encountered
\cite{aichsexl}). However, a simpler method is to note that
because of the Lorentz contraction generated by the boosting, the
gravitational field of the relativistic particle will be
completely confined to the transverse plane, orthogonal to the
instantaneous location of the particle. Hence, before and after
that surface, the space will be vacuum, and the only nontrivial
information about the field will be contained in the junction
conditions at this surface, which separates these vacuum regions.
This enables us to use the cutting and pasting technique of
\cite{thooft,kostas}, which has already been successfully used in
braneworld models \cite{roberto,kallet,kaso}.

So, as in those cases, we start with the vacuum codimension-2
brane solution (\ref{eqn:minkowskipol}), pick a direction on the
brane and switch to lightcone coordinates along it. To encode the
shock wave, we put a relativistic particle along one of the null
lines, say $u=0$, and introduce a discontinuity in the orthogonal
null coordinate $v$ by replacing $dv$ in the metric by $dv -
f(\vec x_\bot, \rho,\phi) \delta(u) du$ \cite{thooft}-\cite{kaso}.
Here $\vec x_\bot$ denotes the spatial dimensions along the brane
which are orthogonal to the direction of motion of the
relativistic source. The shocked metric then becomes
\begin{equation}
ds_D^2= 4dudv - 4\delta(u) fdu^2 + d\vec x^2_\bot + d\rho^2 +
B^2\rho^2d\phi^2 \, . \label{eqn:metric}
\end{equation}
Here $f(\vec x_\bot, \rho,\phi)$ is the shockwave profile, which
only depends on the spatial directions transverse to the motion,
$\vec x_\bot$ along the brane and $\rho,\phi$ away from it.
Further, we add to the brane stress-energy tensor $T^\mu{}_\nu$
the contribution from the momentum of the relativistic particle,
given in terms of the shocked induced brane metric
$g_{D-2\,\mu\nu}$ in (\ref{eqn:metric}) by
\cite{thooft}-\cite{kaso}
\begin{equation}
\tau^\mu{}_\nu = \frac{2p}{\sqrt{g_{D-2}}} \, g_{D-2 \, uv} \,
\delta(u) \, \delta^{(D-4)}(\vec x_\bot) \,  \delta^\mu{}_v \,
\delta^u{}_\nu \, . \label{eqn:sephoton}
\end{equation}
What remains is to substitute (\ref{eqn:metric}) and
(\ref{eqn:sephoton}) into the field equations
(\ref{eqn:compeomstr}), (\ref{eqn:compeomslong}) and work out the
field equation for the shockwave profile $f$. Because
$\tau^\mu{}_\nu$ is traceless, it does not enter in the transverse
field equations (\ref{eqn:compeomstr}), which therefore remain
identical to the vacuum case, and are solved automatically by
(\ref{eqn:metric}) provided that (\ref{eqn:ddimB}) holds. On the
other hand, because the tension term cancels in the longitudinal
equations (\ref{eqn:compeomslong}), as discussed in the text
following Eq. (\ref{eqn:compeomslong}), and $\tau^\mu{}_\mu = 0$,
we find
\be R^\mu{}_{\nu} = \frac{2p}{ M_D^{D-2} B \rho } \, \delta(u) \,
\delta^{(D-4)}(\vec x_\bot) \, \delta(\rho) \, \delta(\phi) \,
\delta^\mu{}_v \, \delta^u{}_\nu \, , \label{eqn:shockeq} \ee
where we have used $g_{D-2 \, uv}/\sqrt{g_{D-2}}= 1$, and $h_{ab}
= {\tt diag}(1, B^2 \rho^2)$ for the metric transverse to the
brane, as per (\ref{eqn:metric}). The only component of
$R^\mu{}_\nu$ which does not vanish trivially is $R^v{}_u$, and
its direct evaluation along the lines of, for example
\cite{kallet,kaso}, yields
\be R^v{}_u =  \, \delta(u) \nabla^2_{D-2} f \, ,
\label{eqn:shockeqcomp} \ee
where $\nabla^2_{D-2} f = \vec \nabla^2_\bot f + \Delta_2 f$ is
the Laplacian defined with respect to the part of the metric
(\ref{eqn:metric}) transverse to the shockwave, spanned by the
coordinates $\vec x_\bot$ and $\rho, \phi$, respectively.
Comparing (\ref{eqn:shockeq}) and (\ref{eqn:shockeqcomp}) yields
the equation for the shockwave profile that we were after:
\be \nabla_{D-2}^2 f = \frac{2p}{ M_D^{D-2} B \rho } \,
\delta^{(D-4)}(\vec x_\bot) \, \delta(\rho) \, \delta(\phi) \, .
\label{eqn:ddimgreensf} \ee
This is the equation for the static potential of a `charge' $p$ at
the origin, on the tip of the cone in $D-2$-dimensional space,
which generates a force with a coupling strength $g \sim
\frac{1}{M_D^{D-2} B}$. It is straightforward to write its
solution, which is\footnote{Since the Laplacian is
$D-2$-dimensional, and the `charge' is at the origin, the solution
must be of the form $f = \frac{\cal Q}{R^{D-4}}$, where $R^2 =
\vec x^2_\bot + \rho^2$. The normalization can be determined from
applying Gauss law to (\ref{eqn:ddimgreensf}), yielding $\int
d\vec S \cdot \vec \nabla f = \frac{2p}{ M_D^{D-2} B}$ and so
${\cal Q}= - \frac{2p} {(D-4) \Omega_{D-3} M_D^{D-2} B}$.}
\be f = - \frac{1} {(D-4) \Omega_{D-3}} \, \frac{2p}{M_D^{D-2} B}
\, \frac{1}{(\vec x^2_\bot + \rho^2)^{\frac{D-4}{2}}} \, ,
\label{eqn:potsoln} \ee
where $\Omega_{D-3} =
\frac{2\pi^{\frac{D-2}{2}}}{\Gamma(\frac{D-2}{2})}$ is the volume
of a unit $S^{D-3}$. Note that the gravitational lightning rod
effect, which we observed in the previous section, remains
manifest in (\ref{eqn:ddimgreensf}). Due to the conical
background, the effective coupling is renormalized from
$1/M_D^{D-2}$ to $1/(M_D^{D-2} B)$, and so it is sensitive to the
brane tension: $g \sim \frac{2\pi}{2\pi M_D^{D-2} - \lambda}$.
Thus the gravitational coupling becomes very strong as the tension
approaches the fundamental scale $M_D$. However, the gravitational
nonlinearities remain under control, being completely suppressed
in the relativistic limit by the boosting of the source. We remark
that the solution (\ref{eqn:potsoln}) is so simple despite the
conical structure of space because the stress-energy source is on
the brane, or equivalently the effective `charge' is on the tip of
the transverse cone. For a source in the bulk off the tip, the
potential of (\ref{eqn:potsoln}) would be more complicated. At
distances short compared to the displacement of the `charge' from
the tip the potential would be the same as in a flat bulk, without
coupling enhancement as the tip is too far to affect it. It would
asymptotically approach (\ref{eqn:potsoln}) as distance increases
\cite{linet}, and would reduce exactly to it as the `charge' is
moved back to the tip of the cone. At any rate, the solution
(\ref{eqn:potsoln}) encapsulates the correct long distance
behavior of the shockwave. We can finally write down the
gravitational field of a relativistic particle zipping along a
codimension-2 brane in $D$-dimensional space time:
\begin{equation}
ds_D^2= 4dudv - \frac{8 p}{(D-4) \Omega_{D-3} M_D^{D-2} B} \,
\frac{\delta(u) \, du^2}{(\vec x^2_\bot + \rho^2)^{\frac{D-4}{2}}}
\,  + d\vec x^2_\bot + d\rho^2 + B^2\rho^2d\phi^2 \, .
\label{eqn:finalmetric}
\end{equation}
In fact this solution looks the same as the higher-dimensional
shockwave in a locally flat spacetime \cite{higherdwaves}, the
only exception being the conical enhancement of the coupling. The
solution (\ref{eqn:finalmetric}) is an exact solution of the field
equations (\ref{eqn:ddimeom}), (\ref{eqn:compeomstr}),
(\ref{eqn:compeomslong}), the brane is thin, with a
$\delta$-function tension as in the vacuum case, but the total
stress-energy tensor on the brane is manifestly {\it not} equal to
pure tension, as can be seen from (see Eqs. (\ref{eqn:ddimeom}),
(\ref{eqn:sephoton}))
\begin{equation}
T^A{}_B = \Bigl( -\lambda \delta^\mu{}_\nu +
\frac{2p}{\sqrt{g_{D-2}}} \, g_{D-2 \, uv} \, \delta(u) \,
\delta^{(D-4)}(\vec x_\bot) \, \delta^\mu{}_v \, \delta^u{}_\nu
\Bigr) \, \delta^A{}_\mu \delta_B{}^\nu \frac{1}{\sqrt{\det
h}}\delta^{(2)}(y) \, . \label{eqn:setotal}
\end{equation}

The solution (\ref{eqn:finalmetric}) remains under control down to
extremely short distances. The reason the shockwave
(\ref{eqn:finalmetric}) evades the results of \cite{cline2,bgsn} is
that in the relativistic limit the gravitational nonlinearities
remain completely under control, as we have discussed above. In
(\ref{eqn:finalmetric}), (\ref{eqn:setotal}) it is clear where the
nonlinearities have `gone': they have been pushed into the metric
discontinuity $\propto \delta(u)$ along the worldline of the source
in (\ref{eqn:finalmetric}), (\ref{eqn:setotal}). While this
$\delta$-function may appear frightful at the first glance, in fact
its divergence is a coordinate artifact that can be easily removed
by a diffeomorphism discussed in \cite{deathpayne,eardley}. Using
(\ref{eqn:metric}) and (\ref{eqn:potsoln}) for notational brevity,
note first that the shockwave is axially symmetric, $\partial_\phi f
= 0$. Further introduce new notation, defining $X^i = (\vec x_\bot,
\rho)$, such that (\ref{eqn:metric}) becomes $ds_D^2 = 4 du dv - 4
\delta(u)fdu^2+ \delta_{ij} dX^i dX^j + B^2 \rho^2 d \phi^2$. Then
define new coordinates $u = \tilde u$, $v = \tilde v + f
\theta(\tilde u) - \tilde u \big(\nabla f\big)^2$, $X^i = \tilde X^i
- 2\tilde u\theta (\tilde u) \tilde
\partial_i f $, where $\theta(\tilde u)$ is the step function,
and new variables are substituted in place of the old ones in the
function $f$ in these transformations. Using $d\big[\theta(\tilde
u)\big]=\delta(\tilde u) d\tilde u$ and $\tilde u\delta(\tilde u)
\equiv 0$, and noting that $\delta(u) f(X) = \delta(\tilde u)
f(\tilde X)$, we can substitute this change of variables in the
metric (\ref{eqn:metric}), (\ref{eqn:potsoln}) to get, after a
straightforward but tedious calculation, the expression
\begin{equation}
ds_D^2 = 4d\tilde u d \tilde v+\Bigl(\delta_{ij} -4\tilde
u\theta(\tilde u) \tilde
\partial_i \tilde
\partial_j f +4\tilde u^2 \tilde \partial_i \tilde \partial_k f \tilde
\partial_j \tilde \partial^k f\Bigr) d\tilde X^id\tilde X^j + B^2 (\tilde \rho
- 2\tilde u\theta (\tilde u) \tilde
\partial_\rho f)^2 d\phi^2  \, , \label{eqn:transformedddimmetric}
\end{equation}
with the form of $f(\tilde X)$ given in (\ref{eqn:potsoln}). This
metric is manifestly well-behaved at $\tilde u =0$.

There is still the singularity at the core of the source, at $\vec
x_\bot = \rho = 0$. Clearly, at any finite distance $|\vec x_\bot|
> 0$ from the source along the brane, there is no bulk divergence
at all. The only singular limit arises in the case of first
approaching $\vec x_\bot = 0$ away from the brane, and then moving
up to it, at the tip of the cone. Although this singularity does
not infect the Ricci curvature, it will show up in the Riemann
tensor, that depends on objects like $\partial_j \partial_k f$.
This however is the usual short distance singularity associated
with any potential source, familiar from electrostatics or
Newtonian gravity. In any case, one expects that at some very
short distance this singularity can be consistently smoothed out
by matter sector effects alone, for example by quantum mechanical
fuzzing up of the source. Therefore, the solution
(\ref{eqn:finalmetric}) is under control as a representation of
the gravitational field of a brane-localized particle. This shows
that brane-localized sources by themselves are not the culprit of
the difficulties with matter-laden thin branes encountered in
\cite{cline2,bgsn}, and subsequently investigated in
\cite{clinvin}-\cite{nasa}. The real cause of these problems is
that gravity is not $4D$ close in, and so it spreads into the bulk
causing strong nonlinear deformations at distances on the order of
the gravitational radius of the energy lump. But this should be
expected all along.

\section{$4D$ Limits}

Having realized what the subtleties with placing matter sources on
thin codimension-2 branes are, it is natural to ask once matter is
included how one can recover $4D$ gravitational force at large
distances. Using a modification of our shockwave geometry
(\ref{eqn:finalmetric}), we will argue here that the recovery of
$4D$ Newton's law may proceed as usual once the scales in the
theory are properly accounted for. We will focus on the case of a
tensional 3-brane in a $6D$ spacetime, although extending the
argument to more dimensions with wrapped branes should be
straightforward. To proceed, let us close the bulk off in some way
at a finite distance from the 3-brane. This could be done in
various ways (see \cite{raman}-\cite{ghesha}, \cite{ahhsw} for
examples). The simplest approach to recovering $4D$ gravity
however is to ignore all the details of compactification, and
merely ask if the correct law at large distances can be so
retrieved.

A simple way to check if this happens is to model the
compactification by imposing some boundary conditions in the bulk,
that remove the `exterior'. A natural trick would be to use
Neumann boundary conditions because they force the gradient of the
potential which would solve (\ref{eqn:ddimgreensf}) to vanish on
some boundary in the radial direction from our 3-brane. This means
that the radial component of the field strength would vanish, and
that the field lines bend around to become parallel with the
3-brane, so that they stop diluting in the transverse directions.
Thus the field strength must switch to the $4D$ law at large
distances. However, implementing this procedure directly on a
stationary field in a compact space requires introducing
unphysical `negative energy' sources on the boundary, by Gauss's
law, and dealing with them, while possible, is unwieldy
\cite{jackson}.

To circumvent these issues, we will instead use periodic boundary
conditions, imposing them by placing images of the brane
throughout the infinite bulk. Although the 3-brane is a cone in
the transverse dimensions, and it is hard to picture a periodic
array of such cones, we will use the fact that the deficit angle
factors into the enhanced gravitational coupling as in
(\ref{eqn:potsoln}), and treat (\ref{eqn:potsoln}) as the
shockwave on a plane. This should be sufficient for our purposes
here. Clearly, a consistent compactification mechanism would have
to be devised to properly account for such short distance issues,
but we can nevertheless test in this way if it can be expected to
reproduce $4D$ gravity at all. So let us imagine that
(\ref{eqn:potsoln}) is promoted into a $2D$ lattice by
translations in the two bulk directions along orthogonal unit
vectors $\vec e_1$ and $\vec e_2$. By linear superposition, the
total shockwave profile of such an array in $D=6$ will be
\be f_{compact} = - \frac{p}{2\pi^2 M_6^4 B} \, \sum_{n_1,n_2 = -
\infty}^{\infty} \frac{1}{\vec x^2_\bot + (\vec \rho - n_1 L \vec
e_1 - n_2 L \vec e_2)^2} \, , \label{eqn:cpotsoln} \ee
where $\vec \rho$ is the bulk component of the radius vector from
the 3-brane at the origin to the point where the potential is
measured, and $L$ is the lattice spacing. We can restrict to
$|\vec \rho| \la L$. At large distances on the brane transverse to
the shock source, $|\vec x_\bot|^2 \gg L^2$, we can approximate
the sum by an integral. Replacing $n_{1,2} \rightarrow y_{1,2}$
(with this normalization $y_k$ are dimensionless), we note that
\be \sum_{n_1,n_2 = - \infty}^{\infty} \frac{1}{\vec x^2_\bot +
(\vec \rho - n_1 L \vec e_1 - n_2 L \vec e_2)^2} \rightarrow
\frac{1}{L^2} \int_{\tt plane}  \frac{d^2 \vec y}{(\vec y - \vec
\rho/L)^2 + \vec x^2_\bot/L^2} \, . \label{eqn:sum} \ee
To evaluate the integral (\ref{eqn:sum}) over an infinite plane,
we shift the origin by a bulk translation $\vec y \rightarrow \vec
y + \vec \rho/L$, without changing the measure of integration, and
then using axial symmetry around the center brane integrate over
the polar angle $\phi$ about it. This yields
\be \frac{1}{L^2} \int_{\tt plane}  \frac{d^2 \vec y}{\vec y^2 +
\vec x^2_\bot/L^2} \,  =  \frac{2\pi}{L^2} \int_0^\infty dy \,
\frac{y}{y^2 + \vec x^2_\bot/L^2} \, , \label{eqn:prefinsum} \ee
The remaining integral is formally infinite because of the
logarithmically divergent contribution of the upper limit of
integration. This infinity is an {\it unphysical} infra-red
divergence arising from the contributions of `charges' infinitely
far away, because their uniform number density far away
overcompensates the potential shutdown with distance. The infinity
is unphysical since the divergent term is pure gauge, and we can
remove it with a diffeomorphism. To do so, we should first
regulate the integral (\ref{eqn:prefinsum}) with a coordinate
space cutoff $\Lambda \gg |\vec x_\bot|/L$, which yields
\be \frac{2\pi}{L^2} \int_0^\infty dy \, \frac{y}{y^2 + \vec
x^2_\bot/L^2} \rightarrow \frac{2\pi}{L^2} \int_0^{\Lambda} dy \,
\frac{y}{y^2 + \vec x^2_\bot/L^2} \, = \frac{\pi}{L^2}
\ln(\frac{{\Lambda}^2 + \vec x_\bot^2/L^2}{{\vec x_\bot^2}/{L^2}})
\, . \label{eqn:finsum} \ee
Next we decompose the logarithm as
\ba \frac{\pi}{L^2} \ln(\frac{{\Lambda}^2 + \vec
x_\bot^2/L^2}{{\vec x_\bot^2}/{L^2}}) &=&  \frac{2\pi}{L^2}
\ln\Lambda - \frac{2 \pi}{L^2} \ln(\frac{|\vec x_\bot|}{L}) +
\frac{\pi}{L^2} \ln(1+ \frac{\vec x_\bot^2}{\Lambda^2 L^2})
\nonumber \\
&=& \frac{2\pi}{L^2} \ln\Lambda - \frac{2 \pi}{L^2}
\ln(\frac{|\vec x_\bot|}{L}) + \frac{\pi}{L^2} \frac{\vec
x_\bot^2}{\Lambda^2 L^2} + \ldots  \, . \label{eqn:logs} \ea
where we have expanded the last logarithm in the top line using
$\Lambda \gg |\vec x_\bot|/L$. Further, we substitute
(\ref{eqn:logs}) into (\ref{eqn:cpotsoln}), and simultaneously
perform the coordinate transformation
\be v \rightarrow v + {\cal A} \, \theta(u) \,
,\label{eqn:vdiffeo} \ee
in the metric (\ref{eqn:metric}), where ${\cal A}$ is a constant
yet to be determined and $\theta(u)$ the step function. Under this
transformation, the shockwave profile changes to
\be f \rightarrow f - {\cal A} \, . \label{eqn:shotrafo} \ee
Then we set ${\cal A} = - \frac{p}{\pi L^2 M_6^4 B} \ln\Lambda$.
This completely cancels the divergent term in the transformed
$f_{compact}$, allowing us to take the limit $\Lambda \rightarrow
\infty$ at will. In this limit, all the cutoff-dependent
polynomial corrections $\propto 1/\Lambda^{2n}$ in
(\ref{eqn:logs}) vanish without a trace. Hence as we promised, the
divergence is completely gauged away, leaving no effect behind.
After introducing the $4D$ Planck mass $M_4^2 = L^2 M_6^{4} B$,
which is precisely the correct Gauss law formula including the
area of the extra-dimensional space, restricted to an elementary
cell of the lattice, we finally find that at large distances along
the brane the shockwave converges to
\be f_{compact} = \frac{p}{\pi M_4^2} \ln(\frac{|\vec x_\bot|}{L})
\, , \label{eqn:grsoln} \ee
The shockwave profile of Eq. (\ref{eqn:grsoln}) is {\it precisely}
the Aichelburg-Sexl $4D$ shockwave solution correctly weighed with
the $4D$ Planck's constant -- just as we have claimed! We see that
the compactification by periodic boundary conditions has
reproduced the $4D$ limit of the solution, with the correctly
normalized $4D$ Planck mass, including the enhancement by the
deficit angle. The exact matching of the numerical coefficients
should not be surprising in spite of the simplicity of the setup,
because of its covariance. Our `compactification prescription'
merely introduced image `charges' which restrict the bulk space to
a finite volume without disturbing the setup. The resulting
periodicity together with the positivity of the potential imply
that there must exist equipotential surfaces around each charge in
the lattice where the potential takes its minimum, and so has
vanishing gradients. This construction is thus effectively
imposing Neumann boundary conditions on the potential minimal
surfaces, without any auxiliary negative `charges'. Based on our
results, we expect that detailed compactification mechanisms with
general matter on codimension-2 branes should work out when the
matter disturbances of the compactification dynamics and the
proper regulators of the matter-laden 3-brane are determined using
all the relevant scales in the problem.

\section{Summary}

In this note we have derived exact black hole and shockwave
solutions localized on a codimension-2 brane in $D$ dimensions,
with gravity governed by the bulk $D$-dimensional Einstein-Hilbert
action. The black hole solutions are higher-dimensional
Schwarzschild geometries with a polar deficit angle, which is
interpreted as a manifestation of the brane's tension that renders
the bulk conical. The solutions are a generalization of the AFV
black hole pierced by a cosmic string in $4D$ \cite{afovi}. They
include a $6D$ black hole on a 3-brane, which can be viewed as an
explicit example of a `small' black hole residing on a 3-brane in
theories with large extra dimensions, with the horizon size
smaller than the size of the extra dimensions, which should be an
interesting arena for explicit calculations of black hole
production and decay rates at the LHC \cite{bhlhc}. Note, that
although our solution (\ref{eqn:bhmetbrn6d}) reduces to $6D$
Schwarzschild when the brane tension is much smaller than the
fundamental scale, when the tension is large a black hole with a
fixed mass, given by the Center-of-Mass energy of the collision in
which it is created, should have a larger radius as dictated by
Eq. (\ref{eqn:horizon}), and hence a greater entropy. This may
improve the semiclassical approximation used to compute black hole
evolution. It would be interesting to test the precise prediction
with the brane tension contributions included, and also seek out
other black hole examples, e.g. with charges and angular momenta.

Our shockwave solutions can be viewed as infinite boost limits of
brane-localized black holes, although we find them by employing
the cut-and-paste tricks of Dray and 't Hooft \cite{thooft}. They
provide an explicit demonstration that gravity really does not
obstruct having localized sources on codimension-2 branes, but
merely obscures their mathematical description because of the
strong nonlinearities at distances comparable to the gravitational
length of the source. For relativistic particles, the boost
restores scaling symmetry pushing the gravitational radius to
zero, and putting nonlinear effects under control. Thus
relativistic particles can be easily described as matter sources
on thin branes, with $\delta$-function stress-energy. The residual
short distance singularities that appear as the distance from the
source goes to zero should be expected to be resolved as usual, by
short distance physics in the core of the source, as for example
the Coulomb singularities of electrostatics which get smeared by
quantum effects. In the case of an infinite locally flat bulk, the
shockwave profiles drop off with distance as $1/r^{D-4}$, i.e. as
$1/r^2$ on a $3$-brane in $6D$, manifestly displaying the
dimensionality of the full spacetime. As a check, we reconsider
the shockwave on a $3$-brane when we close the bulk off by
imposing periodic boundary conditions with a lattice spacing $L$.
In this case we recover the correct logarithmic variation with
distance of the $4D$ Aichelburg-Sexl shockwave at transverse
distances along the brane larger than $L$. These examples support
our view that there exist solutions sourced by stress-energy other
than tension, independently of the internal structure of the brane
and without ever putting higher derivative graviton operators in
the bulk. In general, however, to regulate their mathematical
description correctly, in order to restore the thin brane limit at
large distances, one must account properly for the scales where
the nonlinearities of the gravitational field become important.
While that may be technically involved, it should be possible in
principle.

\vskip1.5cm

{\bf \noindent Acknowledgements}

\smallskip

We thank A.~Lawrence, A.~Rajaraman, L.~Sorbo and J. Terning for
useful discussions. This work was supported in part by the DOE
Grant DE-FG03-91ER40674, in part by the NSF Grant PHY-0332258 and
in part by a Research Innovation Award from the Research
Corporation.

%\pagebreak


\begin{thebibliography}{99}

%\cite{Sundrum:1998ns}
\bibitem{raman}
R.~Sundrum,
%``Compactification for a three-brane universe,''
Phys.\ Rev.\ D {\bf 59} (1999) 085010.
% [arXiv:hep-ph/9807348].
%%CITATION = HEP-PH 9807348;%%

%\cite{Ponton:2000gi}
\bibitem{popo}
A.~Chodos and E.~Poppitz,
%``Warp factors and extended sources in two transverse dimensions,''
Phys.\ Lett.\ B {\bf 471} (1999) 119;
%[arXiv:hep-th/9909199].
%%CITATION = HEP-TH 9909199;%%
E.~Ponton and E.~Poppitz,
%``Gravity localization on string-like defects in codimension two and the
%AdS/CFT correspondence,''
JHEP {\bf 0102} (2001) 042.
%[arXiv:hep-th/0012033].
%%CITATION = HEP-TH 0012033;%%

%\cite{Cohen:1999ia}
\bibitem{cohkap}
A.~G.~Cohen and D.~B.~Kaplan,
%``Solving the hierarchy problem with noncompact extra dimensions,''
Phys.\ Lett.\ B {\bf 470} (1999) 52.
%[arXiv:hep-th/9910132].
%%CITATION = HEP-TH 9910132;%%

%\cite{Gregory:1999gv}
\bibitem{ruth}
R.~Gregory,
%``Nonsingular global string compactifications,''
Phys.\ Rev.\ Lett.\  {\bf 84} (2000) 2564;
% [arXiv:hep-th/9911015].
%%CITATION = HEP-TH 9911015;%%
%\cite{Gregory:2003xf}
%\bibitem{Gregory:2003xf}
% R.~Gregory,
%``Inflating p-branes,''
JHEP {\bf 0306} (2003) 041.
%[arXiv:hep-th/0304262].
%%CITATION = HEP-TH 0304262;%%

%\cite{Chen:2000at}
\bibitem{luty}
J.~W.~Chen, M.~A.~Luty and E.~Ponton,
%``A critical cosmological constant from millimeter extra dimensions,''
JHEP {\bf 0009} (2000) 012.
%[arXiv:hep-th/0003067].
%%CITATION = HEP-TH 0003067;%%

%\cite{Gherghetta:2000qi}
\bibitem{ghesha}
T.~Gherghetta and M.~E.~Shaposhnikov,
%``Localizing gravity on a string-like defect in six dimensions,''
Phys.\ Rev.\ Lett.\  {\bf 85} (2000) 240;
%[arXiv:hep-th/0004014].
%%CITATION = HEP-TH 0004014;%%
%\cite{Gherghetta:2000jf}
%\bibitem{Gherghetta:2000jf}
T.~Gherghetta, E.~Roessl and M.~E.~Shaposhnikov,
%``Living inside a hedgehog: Higher-dimensional solutions that localize
%gravity,''
Phys.\ Lett.\ B {\bf 491} (2000) 353.
%[arXiv:hep-th/0006251].
%%CITATION = HEP-TH 0006251;%%

\bibitem{ahddk}
N.~Arkani-Hamed, S.~Dimopoulos, G.~R.~Dvali and N.~Kaloper,
%``Infinitely large new dimensions,''
Phys.\ Rev.\ Lett.\  {\bf 84} (2000) 586.
% [arXiv:hep-th/9907209].
%%CITATION = HEP-TH 9907209;%%

\bibitem{origami}
N.~Kaloper,
%``Origami world,''
JHEP {\bf 0405} (2004) 061;
% [arXiv:hep-th/0403208].
%%CITATION = HEP-TH 0403208;%%
%``Origami world,''
AIP Conf.\ Proc.\  {\bf 743} (2005) 318.
%%CITATION = APCPC,743,318;%%

%\cite{Carroll:2003db}
\bibitem{cgn}
S.~M.~Carroll and M.~M.~Guica,
%``Sidestepping the cosmological constant with football-shaped extra
%dimensions,''
{\tt arXiv:hep-th/0302067};
%%CITATION = HEP-TH 0302067;%%
%\cite{Navarro:2003vw}
%\bibitem{Navarro:2003vw}
I.~Navarro,
%``Codimension two compactifications and the cosmological constant  problem,''
JCAP {\bf 0309} (2003) 004;
%[arXiv:hep-th/0302129].
%%CITATION = HEP-TH 0302129;%%
%I.~Navarro,
%``Spheres, deficit angles and the cosmological constant,''
Class.\ Quant.\ Grav.\  {\bf 20} (2003) 3603;
%[arXiv:hep-th/0305014].
%%CITATION = HEP-TH 0305014;%%
%\cite{Aghababaie:2003wz}
%\bibitem{Aghababaie:2003wz}
Y.~Aghababaie, C.~P.~Burgess, S.~L.~Parameswaran and F.~Quevedo,
%``Towards a naturally small cosmological constant from branes in 6D
%supergravity,''
Nucl.\ Phys.\ B {\bf 680} (2004) 389;
%[arXiv:hep-th/0304256].
%%CITATION = HEP-TH 0304256;%%
%\cite{Kehagias:2004fb}
A.~Kehagias,
%``A conical tear drop as a vacuum-energy drain for the solution of the
%cosmological constant problem,''
Phys.\ Lett.\ B {\bf 600} (2004) 133;
%[arXiv:hep-th/0406025].
%%CITATION = HEP-TH 0406025;%%
%\cite{Schwindt:2005ns}
%\bibitem{Schwindt:2005ns}
J.~M.~Schwindt and C.~Wetterich,
%``The cosmological constant problem in codimension-two brane models,''
Phys.\ Lett.\ B {\bf 628} (2005) 189.
%[arXiv:hep-th/0508065].
%%CITATION = HEP-TH 0508065;%%

%\cite{Kanti:2001vb}
\bibitem{kmo}
P.~Kanti, R.~Madden and K.~A.~Olive,
%``A 6-D brane world model,''
Phys.\ Rev.\ D {\bf 64} (2001) 044021.
% [arXiv:hep-th/0104177].
%%CITATION = HEP-TH 0104177;%%

%\cite{Nilles:2003km}
\bibitem{nilles}
H.~P.~Nilles, A.~Papazoglou and G.~Tasinato,
%``Selftuning and its footprints,''
Nucl.\ Phys.\ B {\bf 677} (2004) 405;
%  [arXiv:hep-th/0309042].
%%CITATION = HEP-TH 0309042;%%
%\cite{Graesser:2004xv}
%\bibitem{graesser}
M.~L.~Graesser, J.~E.~Kile and P.~Wang,
%``Gravitational perturbations of a six dimensional self-tuning model,''
Phys.\ Rev.\ D {\bf 70} (2004) 024008;
%[arXiv:hep-th/0403074].
%%CITATION = HEP-TH 0403074;%%
%\cite{Garriga:2004tq}
%\bibitem{massimo}
J.~Garriga and M.~Porrati,
%``Football shaped extra dimensions and the absence of self-tuning,''
JHEP {\bf 0408} (2004) 028;
%[arXiv:hep-th/0406158].
%%CITATION = HEP-TH 0406158;%%
%\cite{Redi:2004tm}
M.~Redi,
%``Footballs, conical singularities and the Liouville equation,''
Phys.\ Rev.\ D {\bf 71} (2005) 044006.
%[arXiv:hep-th/0412189].
%%CITATION = HEP-TH 0412189;%%
%%Cited 8 times in SPIRES-HEP

%\cite{Aghababaie:2003ar}
\bibitem{msled}
Y.~Aghababaie {\it et al.},
%``Warped brane worlds in six dimensional supergravity,''
JHEP {\bf 0309} (2003) 037;
%[arXiv:hep-th/0308064].
%%CITATION = HEP-TH 0308064;%%
%\cite{Burgess:2004yq}
%\bibitem{Burgess:2004yq}
C.~P.~Burgess, J.~Matias and F.~Quevedo,
%``MSLED: A minimal supersymmetric large extra dimensions scenario,''
Nucl.\ Phys.\ B {\bf 706} (2005) 71.
%[arXiv:hep-ph/0404135].
%%CITATION = HEP-PH 0404135;%%

%\cite{Cline:2003ak}
\bibitem{cline2}
J.~M.~Cline, J.~Descheneau, M.~Giovannini and J.~Vinet,
%``Cosmology of codimension-two braneworlds,''
JHEP {\bf 0306} (2003) 048.
%[arXiv:hep-th/0304147].
%%CITATION = HEP-TH 0304147;%%

%\cite{Bostock:2003cv}
\bibitem{bgsn}
P.~Bostock, R.~Gregory, I.~Navarro and J.~Santiago,
%``Einstein gravity on the codimension 2 brane?,''
Phys.\ Rev.\ Lett.\  {\bf 92} (2004) 221601.
%[arXiv:hep-th/0311074].
%%CITATION = HEP-TH 0311074;%%

%\cite{Vinet:2004bk}
\bibitem{clinvin}
J.~Vinet and J.~M.~Cline,
%``Can codimension-two branes solve the cosmological constant problem?,''
Phys.\ Rev.\ D {\bf 70} (2004) 083514;
%[arXiv:hep-th/0406141].
%%CITATION = HEP-TH 0406141;%%
%``Codimension-two branes in six-dimensional supergravity and the cosmological
%constant problem,''
Phys.\ Rev.\ D {\bf 71} (2005) 064011.
%[arXiv:hep-th/0501098].
%%CITATION = HEP-TH 0501098;%%

%\cite{deRham:2005ci}
\bibitem{derhato}
C.~de Rham and A.~J.~Tolley,
%``Gravitational waves in a codimension two braneworld,''
{\tt arXiv:hep-th/0511138}.
%%CITATION = HEP-TH 0511138;%%

%\cite{Tolley:2005nu}
\bibitem{tbha}
A.~J.~Tolley, C.~P.~Burgess, D.~Hoover and Y.~Aghababaie,
%``Bulk singularities and the effective cosmological constant for higher
%co-dimension branes,''
{\tt arXiv:hep-th/0512218}.
%%CITATION = HEP-TH 0512218;%%

%\cite{Navarro:2004di}
\bibitem{nasa}
H.~M.~Lee and G.~Tasinato,
%``Cosmology of intersecting brane world models in Gauss-Bonnet gravity,''
JCAP {\bf 0404} (2004) 009;
%[arXiv:hep-th/0401221].
%%CITATION = HEP-TH 0401221;%%
I.~Navarro and J.~Santiago,
%``Gravity on codimension 2 brane worlds,''
JHEP {\bf 0502} (2005) 007;
%[arXiv:hep-th/0411250].
%%CITATION = HEP-TH 0411250;%%
%\cite{Papantonopoulos:2005ma}
%\bibitem{papani}
E.~Papantonopoulos and A.~Papazoglou,
%``Brane-bulk matter relation for a purely conical codimension-2 brane
%world,''
JCAP {\bf 0507} (2005) 004;
%[arXiv:hep-th/0501112].
%%CITATION = HEP-TH 0501112;%%
%\cite{Charmousis:2005ey}
%\bibitem{chaze}
C.~Charmousis and R.~Zegers,
%``Matching conditions for a brane of arbitrary codimension,''
JHEP {\bf 0508} (2005) 075;
%[arXiv:hep-th/0502170].
%%CITATION = HEP-TH 0502170;%%
%\cite{Charmousis:2005ez}
%\bibitem{Charmousis:2005ez}
% C.~Charmousis and R.~Zegers,
%``Einstein gravity on an even codimension brane,''
Phys.\ Rev.\ D {\bf 72} (2005) 064005;
% [arXiv:hep-th/0502171].
%%CITATION = HEP-TH 0502171;%%
%\cite{Kofinas:2005py}
%\bibitem{kofi}
G.~Kofinas,
%``On braneworld cosmologies from six dimensions, and absence thereof,''
{\tt arXiv:hep-th/0506035}.
%%CITATION = HEP-TH 0506035;%%

%\cite{Israel:1976vc}
\bibitem{werner}
W.~Israel,
%``Line Sources In General Relativity,''
Phys.\ Rev.\ D {\bf 15} (1977) 935.
%%CITATION = PHRVA,D15,935;%%

%\cite{Geroch:1987qn}
\bibitem{getra}
R.~Geroch and J.~H.~Traschen,
%``Strings And Other Distributional Sources In General Relativity,''
Phys.\ Rev.\ D {\bf 36} (1987) 1017.
%%CITATION = PHRVA,D36,1017;%%

\bibitem{alex}
A.~Vilenkin and E.~P.~S.~Shellard, {\it Cosmic Strings and Other
Topological Defects}, Cambridge University Press, Cambridge 1994.

%\cite{Aryal:1986sz}
\bibitem{afovi}
M.~Aryal, L.~H.~Ford and A.~Vilenkin,
%``Cosmic Strings And Black Holes,''
Phys.\ Rev.\ D {\bf 34} (1986) 2263.
%%CITATION = PHRVA,D34,2263;%%

%\cite{Achucarro:1995nu}
\bibitem{acgrk}
A.~Achucarro, R.~Gregory and K.~Kuijken,
%``Abelian Higgs hair for black holes,''
Phys.\ Rev.\ D {\bf 52} (1995) 5729.
%[arXiv:gr-qc/9505039].
%%CITATION = GR-QC 9505039;%%

\bibitem{alexcomm} The argument has been stated succintly in
\cite{alex}, page 218, and we quote it here verbatim: {\it ...
Alternatively, one can adopt a more cavalier approach and simply
look for solutions of Einstein's equations with conical
singularities, leaving the mathematical purist to resolve the
question of whether or not they can be interpreted as arising from
distributional sources. ...}

%\cite{Emparan:1999wa}
\bibitem{emhomy}
R.~Emparan, G.~T.~Horowitz and R.~C.~Myers,
%``Exact description of black holes on branes,''
JHEP {\bf 0001} (2000) 007.
%[arXiv:hep-th/9911043].
%%CITATION = HEP-TH 9911043;%%

\bibitem{holograms}
%\cite{Emparan:2002px}
R.~Emparan, A.~Fabbri and N.~Kaloper,
%``Quantum black holes as holograms in AdS braneworlds,''
JHEP {\bf 0208} (2002) 043;
%[arXiv:hep-th/0206155].
%%CITATION = HEP-TH 0206155;%%
T.~Tanaka,
%``Classical black hole evaporation in Randall-Sundrum infinite  braneworld,''
Prog.\ Theor.\ Phys.\ Suppl.\  {\bf 148} (2003) 307.
%[arXiv:gr-qc/0203082].
%%CITATION = GR-QC 0203082;%%

\bibitem{ggi}
%\cite{Galfard:2005va}
C.~Galfard, C.~Germani and A.~Ishibashi,
%``Asymptotically AdS brane black holes,''
{\tt arXiv:hep-th/0512001}.
%%CITATION = HEP-TH 0512001;%%

\bibitem{bhlhc}
P.~C.~Argyres, S.~Dimopoulos and J.~March-Russell,
%``Black holes and sub-millimeter dimensions,''
Phys.\ Lett.\ B {\bf 441} (1998) 96;
% [arXiv:hep-th/9808138].
%%CITATION = HEP-TH 9808138;%%
T.~Banks and W.~Fischler,
%``A model for high energy scattering in quantum gravity,''
{\tt arXiv:hep-th/9906038};
%%CITATION = HEP-TH 9906038;%%
S.~Dimopoulos and G.~Landsberg,
%``Black holes at the LHC,''
Phys.\ Rev.\ Lett.\  {\bf 87} (2001) 161602;
%[arXiv:hep-ph/0106295].
%%CITATION = HEP-PH 0106295;%%
S.~B.~Giddings and S.~D.~Thomas,
%``High energy colliders as black hole factories: The end of short  distance
%physics,''
Phys.\ Rev.\ D {\bf 65} (2002) 056010.
%[arXiv:hep-ph/0106219].
%%CITATION = HEP-PH 0106219;%%

\bibitem{aichsexl}
P.~C.~Aichelburg and R.~U.~Sexl,
%``On The Gravitational Field Of A Massless Particle,''
Gen.\ Rel.\ Grav.\ {\bf 2} (1971) 303.
%%CITATION = GRGVA,2,303;%%

\bibitem{thooft}
T.~Dray and G.~'t Hooft,
%``The Gravitational Shock Wave Of A Massless Particle,''
Nucl.\ Phys.\ B {\bf 253} (1985) 173;
%%CITATION = NUPHA,B253,173;%%
%T.~Dray and G.~'t Hooft,
%``The Gravitational Effect Of Colliding Planar Shells Of Matter,''
Class.\ Quant.\ Grav.\  {\bf 3}  (1986) 825.
%%CITATION = CQGRD,3,825;%%

\bibitem{kostas}
K.~Sfetsos,
%``On gravitational shock waves in curved space-times,''
Nucl.\ Phys.\ B {\bf 436} (1995) 721.
% [arXiv:hep-th/9408169].
%%CITATION = HEP-TH 9408169;%%

\bibitem{roberto}
R.~Emparan,
%``Exact gravitational shockwaves and Planckian scattering on branes,''
Phys.\ Rev.\ D {\bf 64} (2001) 024025.
% [arXiv:hep-th/0104009].
%%CITATION = HEP-TH 0104009;%%

\bibitem{kallet}
N.~Kaloper,
%``Brane-induced gravity's shocks,''
Phys.\ Rev.\ Lett.\  {\bf 94} (2005) 181601;
%[arXiv:hep-th/0501028].
%%CITATION = HEP-TH 0501028;%%
%``Gravitational shock waves and their scattering in brane-induced gravity,''
Phys.\ Rev.\ D {\bf 71} (2005) 086003 [Erratum-ibid: {\bf D71}
(2005) 086003].
%[arXiv:hep-th/0502035].
%%CITATION = HEP-TH 0502035;%%

%\cite{Kaloper:2005wq}
\bibitem{kaso}
N.~Kaloper and L.~Sorbo,
%``Locally localized gravity: The inside story,''
JHEP {\bf 0508} (2005) 070.
%[arXiv:hep-th/0507191].
%%CITATION = HEP-TH 0507191;%%

\bibitem{higherdwaves}
V.~Ferrari, P.~Pendenza and G.~Veneziano,
%``Beamlike Gravitational Waves And Their Geodesics,''
Gen.\ Rel.\ Grav.\  {\bf 20} (1988) 1185;
%%CITATION = GRGVA,20,1185;%%
%\bibitem{devega}
H.~de Vega and N.~Sanchez,
%``Quantum String Scattering In The Aichelburg-Sexl Geometry,''
Nucl.\ Ph.\ B {\bf 317} (1989) 706.
%%CITATION = NUPHA,B317,706;%%

\bibitem{calc}
Perhaps the simplest way to compute the curvature on a cone is to
pick the coordinates such that the transverse metric in
(\ref{eqn:minkowskipol}) is conformally flat, $ds_2 = e^{-2(1-B)
\ln(|\vec z|/\ell)} d\vec z^2$, and to note that because the brane
spacetime is flat, the total curvature is just $R = R_2$. Using
this and eq. (\ref{eqn:ricci}), $R_2 = 2(1-B) e^{2(1-B) \ln(|\vec
z|/\ell)} \vec \nabla_z^2 \ln(|\vec z|/\ell) = 2
\frac{\lambda}{M_D^{D-2}} e^{2(1-B) \ln(|\vec z|/\ell)}
\delta^{(2)}(\vec z)$, or therefore $(1-B) \vec \nabla_z^2
\ln(|\vec z|/\ell) = \frac{\lambda}{M_D^{D-2}} \delta^{(2)}(\vec
z) $. But because $\ln(|\vec z|/\ell)$ is the Euclidean $2D$
Green's function, $\vec \nabla_z^2 \ln(|\vec z|/\ell)= 2\pi
\delta^{(2)}(\vec z)$, so by comparison $ 2\pi (1-B) =
\lambda/M_D^{D-2}$.

\bibitem{rob}
%\cite{Myers:1986un}
R.~C.~Myers and M.~J.~Perry,
%``Black Holes In Higher Dimensional Space-Times,''
Annals Phys.\  {\bf 172} (1986) 304.
%%CITATION = APNYA,172,304;%%

%\cite{Arkani-Hamed:1999dz}
\bibitem{ahhsw}
N.~Arkani-Hamed, L.~J.~Hall, D.~R.~Smith and N.~Weiner,
%``Solving the hierarchy problem with exponentially large dimensions,''
Phys.\ Rev.\ D {\bf 62} (2000) 105002.
%[arXiv:hep-ph/9912453].
%%CITATION = HEP-PH 9912453;%%

%\cite{Linet:1995mq}
\bibitem{linet}
B.~Linet,
%``Euclidean scalar and spinor Green's functions in Rindler space,''
{\tt arXiv:gr-qc/9505033};
%%CITATION = GR-QC 9505033;%%
%\cite{Linet:1995ws}
%B.~Linet,
%``Euclidean thermal spinor Green's function in the space-time of a straight
%cosmic string,''
Class.\ Quant.\ Grav.\  {\bf 13} (1996) 97.
% [arXiv:gr-qc/9506011].
%%CITATION = GR-QC 9506011;%%

%\cite{D'Eath:1992hb}
\bibitem{deathpayne}
P.~D.~D'Eath and P.~N.~Payne,
%``Gravitational radiation in high speed black hole collisions. 1. Perturbation
%treatment of the axisymmetric speed of light collision,''
Phys.\ Rev.\ D {\bf 46} (1992) 658.
%%CITATION = PHRVA,D46,658;%%

%\cite{Eardley:2002re}
\bibitem{eardley}
D.~M.~Eardley and S.~B.~Giddings,
%``Classical black hole production in high-energy collisions,''
Phys.\ Rev.\ D {\bf 66} (2002) 044011.
%[arXiv:gr-qc/0201034].
%%CITATION = GR-QC 0201034;%%

\bibitem{jackson}
J.~D.~Jackson, {\it Classical Electrodynamics}, Second Edition,
John Wiley $\&$ Sons, New York 1975.

\end{thebibliography}
\end{document}